\pdfoutput=1

\documentclass[aps,twocolumn,showpacs,showkeys,amsmath,amssymb,superscriptaddress,nofootinbib,floatfix,prl]{revtex4-1} 

\usepackage{epsfig}

\begin{document} 
\hbadness=10000

\title{Elliptic Flow in Ultrarelativistic Collisions with Polarized Deuterons}

\author{Piotr Bo\.zek}
\email{Piotr.Bozek@fis.agh.edu.pl}
\affiliation{AGH University of Science and Technology, Faculty of Physics and Applied Computer Science, al. Mickiewicza 30, 30-059 Cracow, Poland}

\author{Wojciech Broniowski}
\email{Wojciech.Broniowski@ifj.edu.pl}
\affiliation{Institute of Physics, Jan Kochanowski University, 25-406 Kielce, Poland}
\affiliation{H. Niewodnicza\'nski Institute of Nuclear Physics PAN, 31-342 Cracow, Poland}

\date{ver. 2, 8 October 2018}  

\begin{abstract}
Predictions are made for elliptic flow in collisions of polarized deuterons with a heavy nucleus. 
It is shown that the eccentricity of the initial fireball, evaluated with respect to the deuteron polarization axis perpendicular 
to the beam direction, has a substantial magnitude for collisions of highest multiplicity. Within the Glauber 
approach we obtain $\sim 7\%$ for the deuteron states with spin projection 0, and $\sim -3 \%$ for spin projection $\pm 1$.
We propose to measure   the  elliptic flow coefficient    as the second order harmonic coefficient in the azimuthal 
distribution of  produced charged hadrons with respect to the fixed polarization axis. Collective expansion yields a 
value  of the order of $1\%$ for this quantity, as compared to zero in the absence of polarization and/or collectivity. 
Such a vivid   rotational  symmetry breaking could be  
measured  with the current experimental accuracy of the relativistic heavy-ion experiments. 
The effect has a fundamental significance for understanding the nature 
of dynamics in small systems, as its experimental confirmation would prove the presence of the shape-flow transmutation mechanism, 
typical of hydrodynamic expansion or  rescattering in the later stages of the fireball evolution.
\end{abstract}

\maketitle

The earliest stages of ultrarelativistic light-heavy collisions are an important playground for the strong-interacting dynamics. 
The surprising discovery of the ridge in two-particle correlations, a believed hallmark of collectivity, in p+A collisions~\cite{CMS:2012qk,Abelev:2012ola,Aad:2012gla}, 
followed with d+A~\cite{Adare:2013piz}, and He-A~\cite{Adare:2015ctn}, and even p+p at the highest 
multiplicities of the produced particles \cite{Khachatryan:2010gv},
led to serious considerations  that indeed such small systems may be described by hydrodynamics or transport models, in the same manner as  the large systems formed in A+A
collisions. The early hydrodynamic predictions for harmonic flow in p+A and d+A collisions~\cite{Bozek:2011if} were later confirmed to a surprising accuracy by the experiment~%  
\cite{CMS:2012qk,Abelev:2012ola,Aad:2012gla,Adare:2013piz,Adare:2015ctn}. 
The essential feature of the collective picture applied to these {\em small} systems is
rescattering after the formation of the fireball, which leads to a transmutation, event by event, 
of its transversely deformed shape into the celebrated harmonic flow of 
the finally produced hadrons \cite{Bozek:2011if,Kozlov:2014fqa,Bzdak:2014dia,Bozek:2013uha,Werner:2013tya,Nagle:2013lja}. 
Indeed, this shape-flow transmutation is believed to be one of the key imprints of collectivity of the fireball evolution, 
besides such features as the  mass ordering by collective flow %~\cite{Bozek:2013ska}, 
or the momentum dependence of the femtoscopic radii.

An essential argument in the search for evidence of collective expansion in the final state is the
relation between the geometric deformation of the fireball and the azimuthally asymmetric
flow of emitted hadrons. Whereas in p+A collisions the initial deformation of the fireball originates from fluctuations only,
depending on the model of initial state~\cite{Bozek:2011if,Bzdak:2013zma}, in d+A collisions~\cite{Bozek:2011if}
the elliptic deformation of the fireball is induced by the geometric configuration of the two nucleons in the deuteron. It is dominant
and well constrained by the form of the deuteron wave-function. 
Moreover, in the Glauber model a significant correlation between the event multiplicity and the initial elliptic deformation appears. 
High multiplicity collisions correspond to configuration where the deuteron projectile becomes intrinsically oriented transversely to the beam axis, 
yielding a large number of participant nucleons and a large elliptic deformation~\cite{Bozek:2011if}.
The argument can be generalized to collisions with small projectiles with intrinsic triangular deformation~\cite{Nagle:2013lja,Broniowski:2013dia,Bozek:2014cya}. 
Experimental results from PHENIX Collaboration confirm that the 
hierarchy of elliptic and triangular flows in p+Au, d+Au, and $^3$He+Au collisions follows the  hierarchy of 
the elliptic and triangular deformations of the initial state~\cite{Adare:2013piz,Adare:2015ctn,Aidala:2018mcw}.

At the same time, ongoing efforts are being made within the Color Glass Condensate (CGC) theory to describe
the above-mentioned features of the small systems. 
In this treatment, the dominant part of correlations is generated from the early coherent gluons 
\cite{Dumitru:2008wn,Dusling:2012iga,Kovchegov:2012nd,Dusling:2013oia}.
Naively, one would expect that for configurations corresponding to high multiplicity d+A collisions,  color domains centered around the 
transversely split
projectile neutron and proton contribute independently. Consequently,
the elliptic flow in d+A would be smaller than in p+A collisions, contrary to the experiment. 
However, this argument was recently overturned in~\cite{Mace:2018vwq,Mace:2018yvl}, where the high multiplicity events correspond to
larger saturation scales and to the specific orientation of the deuteron with one of its nucleons behind the other. 

Therefore, the fundamental issue is whether the angular correlations in small systems originate from the initial state dynamics of the gluons or from 
the final state interactions in the fireball. 
Motivated by the dispute, in this Letter we propose an experimental criterion that may probe this issue in a precise and unequivocal manner. Our idea is based on 
the fact that certain light nuclei, such as the deuteron, possess non-zero angular momentum $j$, hence have magnetic moment and thus can be polarized.
In general, if the wave function of the nucleus contains orbital angular momentum $L>0$ components, then the distribution of the nucleons  
in states with good $j_3$ quantum numbers is not spherically symmetric. This allows us to control to some 
degree the ``shape'' of the nuclear distribution in the collision, which is the key trick of our method.

\begin{figure}
\centering
\includegraphics[angle=0,width=0.2 \textwidth]{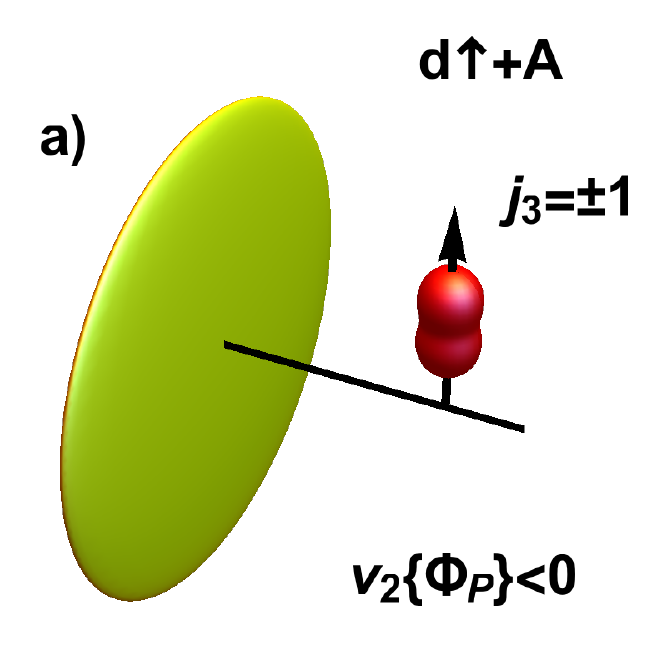} 
\includegraphics[angle=0,width=0.2 \textwidth]{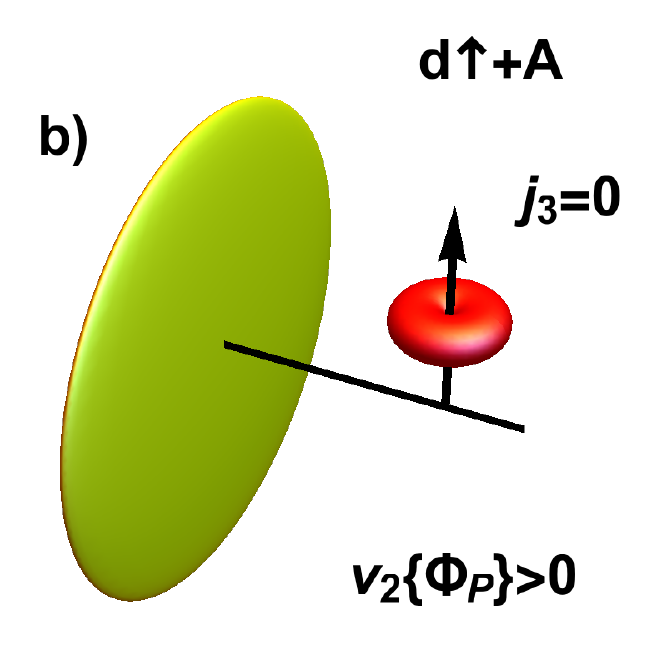} 
\vspace{-4mm}
\caption{A schematic view of the ultrarelativistic d+A collision, where the deuteron is polarized along the axis $\Phi_P$ perpendicular to the beam and has
the spin projection $j_3=\pm 1$ (panel a) or $j_3=0$ (panel b). 
During the collision a fireball is formed, whose orientation in the transverse plane reflects the deformation of the deuteron distribution. 
Via the shape-flow transmutation, the elliptic flow is generated, with the sign as indicated in the figure. \label{fig:polar}}
\end{figure}

The idea is illustrated in Fig.~\ref{fig:polar}. The polarization axis (which is the angular-momentum quantization axis in the rest frame of the deuteron) 
is chosen perpendicularly to the beam, i.e, in the transverse plane. When the deuteron 
angular momentum projection on this axis $j_3=\pm 1$ (panel a), then the distribution of the nucleons at the reaction is prolate. Upon collisions with the 
nucleons from the big nucleus (the flattened disk in the figure), the formed fireball is also prolate in the transverse plane, 
simply reflecting the distribution in the deuteron. Then, if collectivity 
takes over in the proceeding evolution, the elliptic flow coefficient evaluated with respect to the polarization axis is negative, $v_2\{\Phi_P\}<0$.
For the state $j_3=0$ (panel b), the situation is opposite, with now an oblate shape and $v_2\{\Phi_P\}>0$. Of course, the crucial question is the 
magnitude of the effect. We show that in fact it is within the experimental resolution of the current experiments, even if realistic (not 100\%) polarization 
of the deuteron is achieved.

The basic measures in the collective flow analysis are 
the eccentricity vector corresponding to the azimuthal asymmetry of the 
 initial density $f(\vec{\rho})$ in the transverse plane $(\rho,\alpha)$,  
\begin{eqnarray}
 \vec{\epsilon}_n  &=& \epsilon^x_n + i  \epsilon^y_n  
 = - \frac{\int \rho d\rho d\alpha e^{i n \alpha}\rho^n f(\vec{\rho})}{ \int \rho d\rho d\alpha \rho^n f(\vec{\rho})} \ , \label{eq:eps} 
\end{eqnarray}
and the flow vector determined from the azimuthal distribution in the event $dN^{\rm ev}/d\phi$ 
\begin{eqnarray}
 \vec{v}_n  &=& v^x_n + i  v^y_n  
 =  \frac{\int  d\phi e^{i n \phi} \frac{dN^{\rm ev}}{d\phi}}{\int  d\phi \frac{dN^{\rm ev}}{d\phi}}, \label{eq:v} 
\end{eqnarray}
with $n$ denoting the Fourier rank. The essential feature of collective evolution % involving hydrodynamics or transport approaches 
is that the 
eccentricity and flow vectors are to a good approximation proportional to each other event by event. In particular, for the considered 
elliptic flow
\begin{equation}
\vec{v}_2 \simeq k \vec{\epsilon}_2, \label{eq:prop}
\end{equation}
where the coefficient $k \sim 0.2$ for the considered small systems~\cite{Nagle:2013lja}. %The relation between the deformation of the initial source and the final flow (\ref{eq:prop}) holds event by event.
 For collisions with unpolarized deuterons the orientation of 
the eccentricity $\vec{\epsilon}_2$ and flow $\vec{v}_2$ vectors is random.
 The azimuthal distribution  in an event $dN^{\rm ev}/d\phi$ cannot be extracted 
from the observed particles with  finite multiplicity.
 Flow coefficients  can be estimated from multiparticle 
distributions, as discussed below. 
On the other hand, collisions with polarized beams give control on the orientation of the deuteron deformation  using 
 the eccentricity and flow  vectors projected on the {\it fixed}  polarization axis $\Phi_P$, 
\begin{eqnarray}
\epsilon_n\{\Phi_P\} & \equiv & \epsilon^x_n  \cos \Phi_P +    \epsilon^y_n \sin \Phi_P, \nonumber \\
 v_n\{\Phi_P\} & \equiv & v^x_n  \cos \Phi_P + v^y_n \sin \Phi_P. \label{eq:ep}
\end{eqnarray} 
Clearly, the proportionality of Eq.~(\ref{eq:prop}) 
holds also for the projected quantities of Eq.~(\ref{eq:ep}).

The deuteron is a $j^P=1^+$ state, with a dominant ${}^3S_{1}$-wave component and a 
few percent ${}^3D_{1}$-wave admixture. With these two components, the wave function with $j_3$ projection of the total angular momentum $j$ can be
written as
\begin{eqnarray}
| \Psi(r;j_3) \rangle &=& U(r)|j=1,j_3,L=0,S=1 \rangle  \nonumber \\ &+&  V(r)|j=1,j_3,L=2,S=1 \rangle,  
\end{eqnarray}
where $r$ in the relative radial coordinate, and $U(r)$ and $V(r)$ are the $S$ and $D$ radial functions, respectively.  
Explicitly, with the Clebsch-Gordan decomposition into states $|LL_3\rangle |SS_3\rangle$, 
\begin{eqnarray}
&& | \Psi(r;1)\rangle = U(r) |00 \rangle |11 \rangle \label{eq:02} \\ 
&& ~+ V(r)  \Big [ \sqrt{\tfrac{3}{5}}  |22 \rangle |1-\!\!1 \rangle -  \sqrt{\tfrac{3}{10}}  |21 \rangle |10 \rangle + \sqrt{\tfrac{1}{10}}  |20 \rangle |11 \rangle \Big ] , \nonumber \\
&& | \Psi(r;0)\rangle = U(r) |00 \rangle |10 \rangle \nonumber  \\ 
&& ~+ V(r)  \Big [ \sqrt{\tfrac{3}{10}}  |21 \rangle |1-\!\!1 \rangle -  \sqrt{\tfrac{2}{5}}  |20 \rangle |10 \rangle + \sqrt{\tfrac{3}{10}}  |2-\!\!1 \rangle |11 \rangle \Big ] . \nonumber
\end{eqnarray}
Further, orthonormality of the spin parts yields the following expressions for the moduli squared of the wave functions: 
\begin{eqnarray}
&& | \Psi(r,\theta,\phi;\pm 1)|^2 = \frac{1}{16\pi}  \left [ 4 U(r)^2 - \right . \label{eq:dens} \\ 
&& ~~\left . 2 \sqrt{2} \left(1-3 \cos ^2(\theta ) \right) U(r) V(r)+\left(5-3 \cos ^2(\theta )\right) V(r)^2 \right ], \nonumber \\
&& | \Psi(r,\theta,\phi;0)|^2 = \frac{1}{8\pi}  \left [ 2 U(r)^2 + \right . \nonumber \\ 
&& ~~\left . 2 \sqrt{2} \left(1-3 \cos ^2(\theta )\right) U(r) V(r)+\left(1+3 \cos ^2(\theta )\right) V(r)^2 \right ],  \nonumber
\end{eqnarray}
with 
%\begin{eqnarray}
$\sum_{j_3} | \Psi(r,\theta,\phi;j_3)|^2=\frac{3}{4\pi}[U(r)^2+V(r)^2]$. % \label{eq:rot}
%\end{eqnarray}

We are being so explicit to point out several features. First, the interference term between the spin $|11\rangle$ 
components in the wave functions of Eq.~(\ref{eq:02}), giving the terms proportional to $U(r) V(r)$ in Eq.~(\ref{eq:dens}), is crucial for a significant polar angle dependence. 
This is because $V(r)^2 \ll U(r)^2$ and the terms proportional to $V(r)^2$ are negligible. 
Second, we note that the densities of Eq.~(\ref{eq:dens}) are prolate for $j_3=\pm1$, and oblate for 
$j_3=0$ (cf. Fig.~\ref{fig:polar}).

\begin{figure}
\centering
\includegraphics[angle=0,width=0.34 \textwidth]{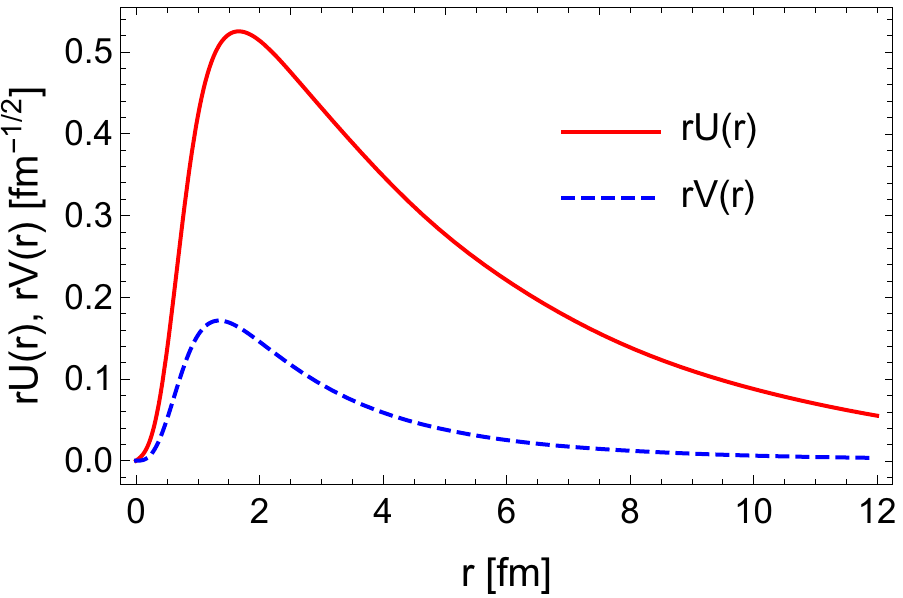}
\vspace{-2mm}
\caption{Radial wave functions of the $S$-wave, $U(r)$, and $D$-wave, $V(r)$, components of the deuteron, multiplied by the relative radius 
$r$, taken from the parametrization provided in~\cite{Zhaba:2015yxq} for  Reid93 nucleon-nucleon potential.  \label{fig:reid93}}
\end{figure}

There are many parameterizations of the deuteron radial wave functions in the literature~\cite{Zhaba:2015yxq}, yielding similar results. 
Here we use the wave functions obtained from Reid93 nucleon-nucleon potential, shown in Fig.~\ref{fig:reid93}. 
In this parametrization, the weight of the $D$-wave part in the probability distribution is
$\int_0^\infty V(r)^2 r^2 dr = 5.7\%$, clearly exhibiting the strong $S$-wave dominance.
It is interesting to examine the ellipticity of the distribution of Eq.~(\ref{eq:dens}), defined in analogy to Eq.~(\ref{eq:eps},\ref{eq:ep})
for $n=2$ with $f(\vec{\rho})$ replaced with the modulus squared of the deuteron wave function. We get
\begin{eqnarray}
&& \epsilon_2^{|\Psi|^2_{j_3=0}}\{\Phi_P\} = \label{eq:epsi} \\
&&  \frac{\int  d^3r \, r^2 \{ \tfrac{2\sqrt{2}}{5} U(r)V(r)\!-\!\tfrac{1}{5}V(r)^2 \} }
 {\int d^3r \, r^2 \{ \tfrac{2}{3} U(r)^2\!-\!\tfrac{2\sqrt{2}}{15} U(r)V(r)\!+\!\tfrac{11}{15} V(r)^2  \} } \simeq 0.11, \nonumber \\
&& \epsilon_2^{|\Psi|^2_{j_3 = \pm 1}}\{\Phi_P\} =\nonumber \\
&& \frac{\int \! d^3r \, r^2 \{- \tfrac{\sqrt{2}}{5} U(r)V(r)\!+\!\tfrac{1}{10}V(r)^2 \} }
 {\int \!\! d^3r \, r^2 \{ \tfrac{2}{3} U(r)^2\!+\!\tfrac{\sqrt{2}}{15} U(r)V(r)\!+\!\tfrac{19}{30} V(r)^2  \} } \simeq -0.05 \nonumber
\end{eqnarray}
(projection of the distribution on the transverse plane provides here an extra dimension in the integration compared to Eq.~(\ref{eq:eps})).
As already mentioned, the relatively large values of these eccentricities are caused by the interference term with $U(r) V(r)$. 
We note that approximately
%\begin{eqnarray}
$ \epsilon_2^{|\Psi|^2_{j_3 = \pm 1}}\{\Phi_P\} \simeq -\frac{1}{2} \epsilon_2^{|\Psi|^2_{j_3=0}}\{\Phi_P\}$. 
%\label{eq:approx}
%\end{eqnarray}

In the Glauber approach, the nucleons from the deuteron interact (incoherently) with the nucleons of the target. The reaction,
shorter than any nuclear time scale
due to a huge Lorentz contraction factor, causes the reduction of the wave functions of both the projectile and the target, 
with nucleons acquiring positions in the transverse plane. The 
eccentricity of the deuteron wave function discussed above is thus reflected in the distribution of its nucleons. Upon collisions with the nucleons of the target, 
a corresponding eccentricity of the fireball is generated. It
can be quantified with  Eq.~(\ref{eq:ep}), where $f(\vec{\rho})$ is
the distribution of entropy in a given event, and averaging over events in $\epsilon_2\{\Phi_P\}$ is understood.
\begin{figure}
\begin{center}
\includegraphics[angle=0,width=0.42\textwidth]{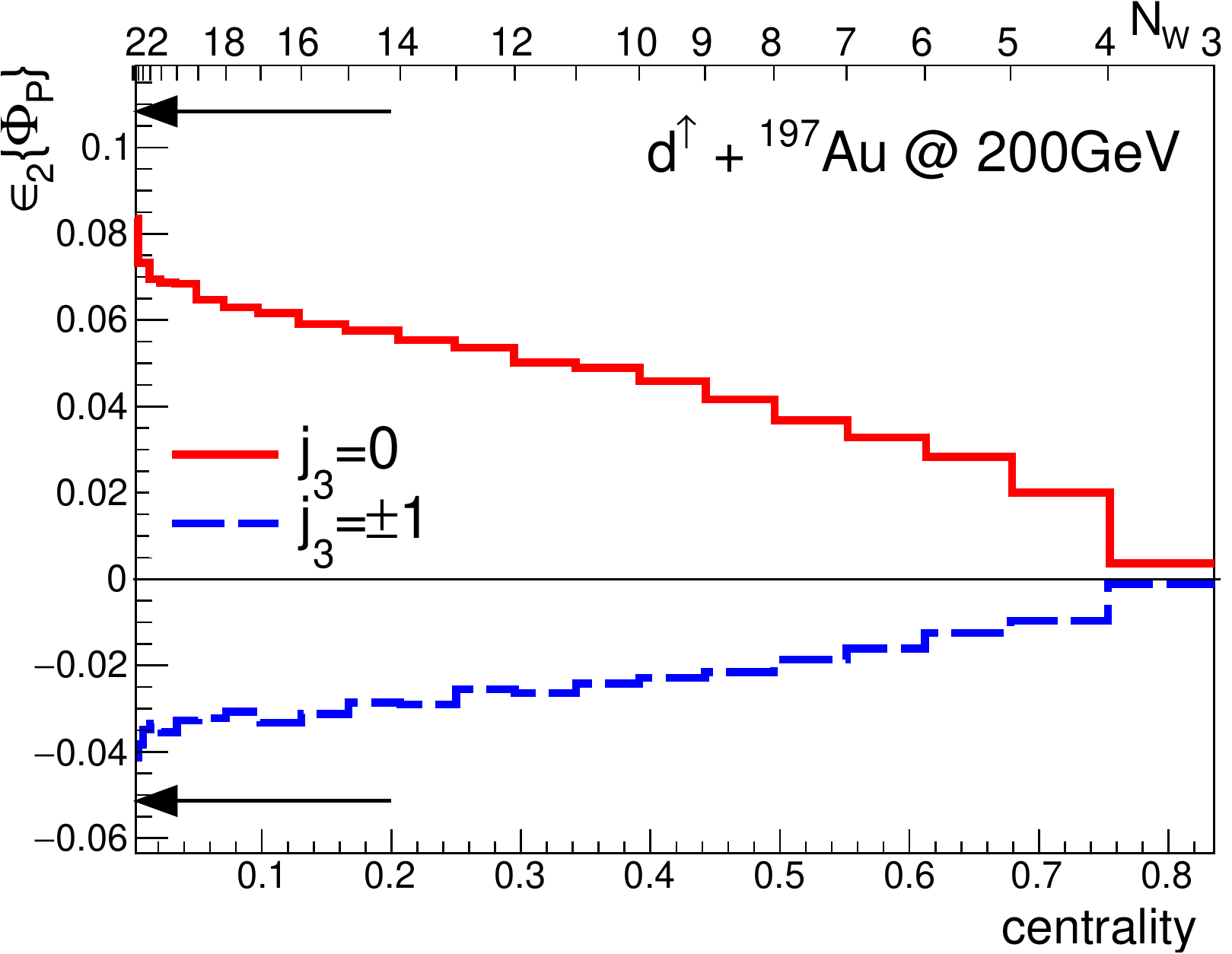}
\end{center}
\vspace{-6mm}
\caption{Ellipticities of the fireball formed in  polarized d+Au collisions at the energy of 
$\sqrt{s_{NN}}=200$~GeV. The lower coordinate axis shows the centrality as defined via the produced entropy $S$.
The top coordinate axis shows the corresponding number of the wounded nucleons. 
The arrows indicate the ellipticities of the modulus squared of the deuteron wave function of Eq.~(\ref{eq:epsi}).  \label{fig:e2P}}
\end{figure}
The discussed effect is generic and appears in any variant of the Glauber model.
In our study, we use the wounded nucleon model~\cite{Bialas:1976ed} with a binary collisions admixture~\cite{Kharzeev:2000ph}, as implemented in
the Glauber Monte Carlo code {\tt GLISSANDO}~\cite{Rybczynski:2013yba}.
The production of the initial entropy is proportional to 
%\begin{eqnarray}
$S= {\rm const} \left ( N_{\rm W}/2 + a N_{\rm bin} \right )$, 
%\label{eq:S}
% \end{eqnarray}
with the parameter $a=0.145$, whereas $N_{\rm W}$ and   $N_{\rm bin}$
are the numbers of the wounded nucleons and binary collisions, respectively. 
The deposition of the entropy at the NN collision point in the transverse plane  is 
smeared with a Gaussian of width $0.4$~fm. 
The  results of the simulations for  $\epsilon_2\{\Phi_P\}$ of the fireball are shown in Fig.~\ref{fig:e2P}. The centrality of the collision
is defined via quantiles of the distribution of the initial entropy  $S$. 
%of Eq.~(\ref{eq:S}). 
For convenience, we also show the corresponding number of the
the wounded nucleons, $N_{\rm W}$, on the top coordinate axis. 
We note that for the most central collisions (large $N_{\rm W}$), the ellipticities 
of the fireball are reduced by $\sim 30\%$ compared to the ellipticities of the distributions of the polarized deuteron of Eq.~(\ref{eq:epsi}), 
indicated with arrows. This reduction is caused by the contribution from the Au nucleons, whose positions fluctuate. 
The effect is stronger as $N_{\rm W}$ decreases, with $\epsilon_2\{\Phi_P\}$ dropping to zero for peripheral collisions. 
We note that the approximate relation $\sum_{j_3} \epsilon_2^{j_3}\{\Phi_P\} \simeq 0$ is satisfied, in accordance to 
the corresponding relation for the eccentricities of the wave functions.
Importantly, the size of $\epsilon_2\{\Phi_P\}$ is at the level of a few percent, which is a sizable value.
According to Eq.~(\ref{eq:prop}), the corresponding values of $v_2\{\Phi_P\}$ for the reaction of Fig.~\ref{fig:e2P} 
are expected to be of the order of 1\% for the most central collisions, compared to zero in the absence of polarization and/or collective evolution. 

The experimental observation of the proposed effect requires the use of polarized beams or targets \cite{Mane:2005xh,Alekseev:2003sk}.
For particles of angular momentum $1$, the vector polarization is $P_z=n(1)-n(-1)$, and the tensor polarization, relevant for our 
proposal, is 
$P_{zz}=n(1)+n(-1)-2 n(0)$, where $n(j_3)$ denotes the fraction of states with angular momentum projection $j_3$. 
Since in our case the magnitude of the eccentricity of the fireball is about twice as large  for collisions with deuteron in $j_3=0$ state than in $j_3=\pm 1$ state,
the total predicted  elliptic flow with respect to the 
polarization axis for (partially) polarized deuterons is 
\begin{eqnarray}
v_2\{\Phi_P\} \simeq k\,  \epsilon_2^{j_3=\pm1}\{\Phi_P\} P_{zz}. \label{eq:polex}
\end{eqnarray}
It is maximal and positive for $P_{zz}=-2$, reaching about 1.5\%, and minimal and negative for $P_{zz}=1$, reaching about $-0.75$\% for most central collisions.
For the deuteron, experimentally achievable polarization is within the range $-1.5 \lesssim P_{zz} \lesssim 0.7$~\cite{Sato:1996te,Savin:2014sva}, 
which according Eq.~(\ref{eq:polex}) yields of $-0.5\% \lesssim v_2\{\Phi_P\} \lesssim 1\%$. With the present accuracy of  elliptic flow 
measurements, this size of effect could be measured. 

\begin{figure}
\centering
\includegraphics[angle=0,width=0.42 \textwidth]{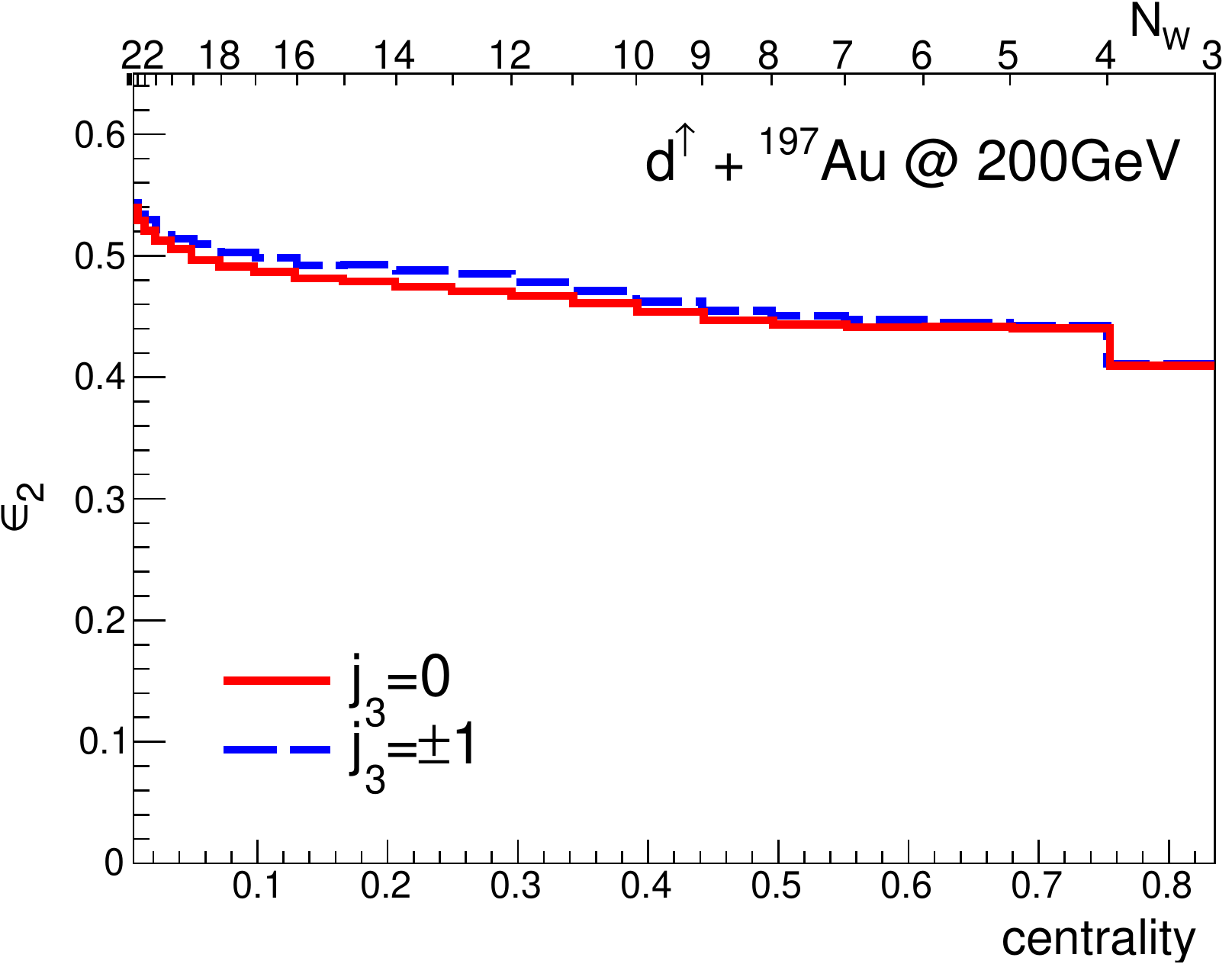}
\vspace{-3mm}
\caption{Same as in Fig.~\ref{fig:e2P} but for the participant-plane ellipticity $\epsilon_2$. 
It is dominated by fluctuations and the relative splitting effect between the $j_3=0$ and $j_3=\pm 1$ cases is small. \label{fig:e2}}
\end{figure}

Next, we discuss the difference between our proposal and the standard estimates of the elliptic flow, used in most analyses up to now. 
There, the orientation of the eccentricity (\ref{eq:eps}) and of the flow asymmetry (\ref{eq:v})  fluctuates randomly. 
To extract the $v_2$ coefficient,
methods involving two- or (more-) particle correlations must be used. 
 The two-particle cumulant estimate~\cite{Borghini:2001vi} ($v_2\{2\}$) is based on the the two-particle distribution
\begin{equation}
\frac{dN}{d\phi_1 d\phi_2} \propto 1+ 2v_2\{2\}^2 \cos\left[ 2(\phi_1-\phi_2) \right] +\dots .
\label{eq:azpol2}
\end{equation}

On the other hand, the elliptic flow  projected on the polarization axis 
$v_2\{\Phi_P \}$  can be measured  using  the one-particle distribution, 
which is deformed relative to the 
known polarization direction $\Phi_P$, 
\begin{equation}
\frac{dN}{d\phi} \propto 1+ 2v_2\{ \Phi_P \} \cos\left[ 2(\phi-\Phi_P) \right] +\dots .
\label{eq:azpol}
\end{equation}
This has important advantages from the experimental point of view.
The cumulant methods estimate higher powers of the small flow coefficient, hence a larger statistics is required \cite{Borghini:2001vi} 
as compared to the measurement 
of $v_2\{\Phi_P\}$,
%the flow coefficient with respect to the polarization axis from the 
%one-particle distribution (\ref{eq:azpol}), 
especially for collisions with small multiplicity. Although the projection on the polarization axis 
reduces somewhat the flow coefficient  $v_2\{\Phi_P \}$ as compared  to $v_2\{2\}$, the elliptic flow in d+A collisions is small and
we have $v_2\{\Phi_P \} > v_2\{2\}^2$. %, pointing out a much larger effect in Eq.~(\ref{eq:azpol})  compared to Eq. ~(\ref{eq:azpol2}). 
Secondly, it is well known that measurements using cumulants of the correlation function
contain systematic uncertainties from non-flow effects, e.g. 
%. For example, the two-particle 
% distribution function  may contain correlations from 
from resonance decays or jets.
% besides the flow correlations given  in Eq. (\ref{eq:azpol2}). 
On the other hand, the elliptic  flow  with respect to the polarization axis 
simply measures the 
azimuthal asymmetry of the final hadrons from the one-particle  distribution.
Thirdly, in collisions with polarized deuterons the azimuthal asymmetry 
of emitted hard probes, i.e.,  jets, photons or heavy flavor mesons, can be measured with respect 
to the polarization axis. In standard flow analyses the azimuthal asymmetry 
of hard probes is defined from the correlation with other (soft momentum) particles. 
Finally, interferometry correlations for same-charge pion pairs can be determined
for the pairs emitted in the directions along or perpendicular
to the polarization axis. That way a  possible azimuthal asymmetry of the pion emission 
sources in the fireball could be observed.

The participant-plane ellipticity, $\epsilon_2=|\vec{\epsilon_2}|$, 
for d+Au collisions simulated with {\tt GLISSANDO} is shown in 
Fig.~\ref{fig:e2}. In this case
the eccentricity is dominated by fluctuations and the relative splitting effect between 
the $j_3=0$ and $j_3=\pm 1$ cases is tiny and could not be unraveled 
with present model and experimental uncertainties. This illustrates the  advantages of our proposal discussed above. 
%that allows to experimentally tune  the {\it orientation} of the polarized projectile.
The measurement of a small but nonzero elliptic flow  with respect to the polarization axis of Eq.~(\ref{eq:ep}) is essential 
for the verification of the effect of the shape-flow transmutation in small systems.

We present calculations for the BNL Relativistic Heavy-Ion Collider  energies of $200$~GeV, but the results are similar for other
energies, such as at the CERN Large Hadron Collider. Our predictions could 
also be tested at lower energies where it is easier to deliver a polarized 
deuteron beam, or even in experiments with heavy ion beams colliding with a fixed polarized target, such as possible in the 
NA61 setup~\cite{Abgrall:2014xwa} 
at the CERN  Super Proton Synchrotron, or in the planned LHCb fixed target run
(SMOG)~\cite{Barschel:2014iua,Aaij:2014ida}.
We note that the effect discussed here for the deuteron occurs for other $j\ge 1$ nuclei as well. 
The constraint  $j\ge 1$ originates from the angular-momentum algebra:
the numerator of the eccentricity in Eq.~(\ref{eq:ep}) is
a tensor operator of rank two, hence (up to tiny corrections from the denominator) the eccentricity has 
non-vanishing diagonal matrix elements between states of  $j\ge 1$.
Thus, we expect a similar size and behavior of $\epsilon\{\Phi_P\}$ for
such nuclei as ${}^{7}$Li or ${}^{7,9}$Be, which have $j=3/2$, and no effect for  ${}^{3}$H or ${}^{3}$He, which 
are $j=1/2$ states. A rough measure of the admixture of $L>0$ states in the wave function 
is the mismatch of the total magnetic moment from the sum of magnetic moments of the nucleonic spins.
For the deuteron, the mismatch is 3\%, whereas  for ${}^7$Li -- 14\%, and for ${}^9$Be -- 60\%, thus we expect 
the effect to be stronger there. In lithium or beryllium nuclei, a strong intrinsic deformation is linked to their cluster structure. 
%~\cite{Rybczynski:2017nrx}. 
Precise estimates %of the proposed effect for these nuclei 
are left for a separate study. 

We thank Adam Bdzak for  clarifying and inspiring discussions on alternative CGC and hydrodynamic
explanations of collectivity in small systems, and Janusz Chwastowski for discussions concerning polarization.
We are grateful for the hospitality of  Centro  de Ciencas de Benasque and Munich Institute for Astro- and Particle Physics (MIAPP), where this research was done.
Supported by the  Polish National
Science Centre grants 2015/17/B/ST2/00101 (PB) and 2015/19/B/ST2/00937 (WB), 
by the COST Action CA15213 ``THOR'', 
and by MIAPP DFG cluster of excellence ``Origin and Structure of the Universe".

\bibliography{../../hydr}

\begin{thebibliography}{34}%
\makeatletter
\providecommand \@ifxundefined [1]{%
 \@ifx{#1\undefined}
}%
\providecommand \@ifnum [1]{%
 \ifnum #1\expandafter \@firstoftwo
 \else \expandafter \@secondoftwo
 \fi
}%
\providecommand \@ifx [1]{%
 \ifx #1\expandafter \@firstoftwo
 \else \expandafter \@secondoftwo
 \fi
}%
\providecommand \natexlab [1]{#1}%
\providecommand \enquote  [1]{``#1''}%
\providecommand \bibnamefont  [1]{#1}%
\providecommand \bibfnamefont [1]{#1}%
\providecommand \citenamefont [1]{#1}%
\providecommand \href@noop [0]{\@secondoftwo}%
\providecommand \href [0]{\begingroup \@sanitize@url \@href}%
\providecommand \@href[1]{\@@startlink{#1}\@@href}%
\providecommand \@@href[1]{\endgroup#1\@@endlink}%
\providecommand \@sanitize@url [0]{\catcode `\\12\catcode `\$12\catcode
  `\&12\catcode `\#12\catcode `\^12\catcode `\_12\catcode `\%12\relax}%
\providecommand \@@startlink[1]{}%
\providecommand \@@endlink[0]{}%
\providecommand \url  [0]{\begingroup\@sanitize@url \@url }%
\providecommand \@url [1]{\endgroup\@href {#1}{\urlprefix }}%
\providecommand \urlprefix  [0]{URL }%
\providecommand \Eprint [0]{\href }%
\providecommand \doibase [0]{http://dx.doi.org/}%
\providecommand \selectlanguage [0]{\@gobble}%
\providecommand \bibinfo  [0]{\@secondoftwo}%
\providecommand \bibfield  [0]{\@secondoftwo}%
\providecommand \translation [1]{[#1]}%
\providecommand \BibitemOpen [0]{}%
\providecommand \bibitemStop [0]{}%
\providecommand \bibitemNoStop [0]{.\EOS\space}%
\providecommand \EOS [0]{\spacefactor3000\relax}%
\providecommand \BibitemShut  [1]{\csname bibitem#1\endcsname}%
\let\auto@bib@innerbib\@empty
%</preamble>
\bibitem [{\citenamefont {Chatrchyan}\ \emph {et~al.}(2013)\citenamefont
  {Chatrchyan} \emph {et~al.}}]{CMS:2012qk}%
  \BibitemOpen
  \bibfield  {author} {\bibinfo {author} {\bibfnamefont {S.}~\bibnamefont
  {Chatrchyan}} \emph {et~al.} (\bibinfo {collaboration} {CMS Collaboration}),\
  }\href@noop {} {\bibfield  {journal} {\bibinfo  {journal} {Phys. Lett.}\
  }\textbf {\bibinfo {volume} {B718}},\ \bibinfo {pages} {795} (\bibinfo {year}
  {2013})},\ \Eprint {http://arxiv.org/abs/1210.5482} {arXiv:1210.5482
  [nucl-ex]} \BibitemShut {NoStop}%
%%CITATION = ARXIV:1210.5482;%%
\bibitem [{\citenamefont {Abelev}\ \emph {et~al.}(2013)\citenamefont {Abelev}
  \emph {et~al.}}]{Abelev:2012ola}%
  \BibitemOpen
  \bibfield  {author} {\bibinfo {author} {\bibfnamefont {B.}~\bibnamefont
  {Abelev}} \emph {et~al.} (\bibinfo {collaboration} {ALICE Collaboration}),\
  }\href {\doibase 10.1016/j.physletb.2013.01.012} {\bibfield  {journal}
  {\bibinfo  {journal} {Phys. Lett.}\ }\textbf {\bibinfo {volume} {B719}},\
  \bibinfo {pages} {29} (\bibinfo {year} {2013})},\ \Eprint
  {http://arxiv.org/abs/1212.2001} {arXiv:1212.2001 [nucl-ex]} \BibitemShut
  {NoStop}%
%%CITATION = ARXIV:1212.2001;%%
\bibitem [{\citenamefont {Aad}\ \emph {et~al.}(2013)\citenamefont {Aad} \emph
  {et~al.}}]{Aad:2012gla}%
  \BibitemOpen
  \bibfield  {author} {\bibinfo {author} {\bibfnamefont {G.}~\bibnamefont
  {Aad}} \emph {et~al.} (\bibinfo {collaboration} {ATLAS Collaboration}),\
  }\href {\doibase 10.1103/PhysRevLett.110.182302} {\bibfield  {journal}
  {\bibinfo  {journal} {Phys. Rev. Lett.}\ }\textbf {\bibinfo {volume} {110}},\
  \bibinfo {pages} {182302} (\bibinfo {year} {2013})},\ \Eprint
  {http://arxiv.org/abs/1212.5198} {arXiv:1212.5198 [hep-ex]} \BibitemShut
  {NoStop}%
%%CITATION = ARXIV:1212.5198;%%
\bibitem [{\citenamefont {Adare}\ \emph {et~al.}(2013)\citenamefont {Adare}
  \emph {et~al.}}]{Adare:2013piz}%
  \BibitemOpen
  \bibfield  {author} {\bibinfo {author} {\bibfnamefont {A.}~\bibnamefont
  {Adare}} \emph {et~al.} (\bibinfo {collaboration} {PHENIX Collaboration}),\
  }\href {\doibase 10.1103/PhysRevLett.111.212301} {\bibfield  {journal}
  {\bibinfo  {journal} {Phys. Rev. Lett.}\ }\textbf {\bibinfo {volume} {111}},\
  \bibinfo {pages} {212301} (\bibinfo {year} {2013})},\ \Eprint
  {http://arxiv.org/abs/1303.1794} {arXiv:1303.1794 [nucl-ex]} \BibitemShut
  {NoStop}%
%%CITATION = ARXIV:1303.1794;%%
\bibitem [{\citenamefont {Adare}\ \emph {et~al.}(2015)\citenamefont {Adare}
  \emph {et~al.}}]{Adare:2015ctn}%
  \BibitemOpen
  \bibfield  {author} {\bibinfo {author} {\bibfnamefont {A.}~\bibnamefont
  {Adare}} \emph {et~al.} (\bibinfo {collaboration} {PHENIX Collaboration}),\
  }\href {\doibase 10.1103/PhysRevLett.115.142301} {\bibfield  {journal}
  {\bibinfo  {journal} {Phys. Rev. Lett.}\ }\textbf {\bibinfo {volume} {115}},\
  \bibinfo {pages} {142301} (\bibinfo {year} {2015})},\ \Eprint
  {http://arxiv.org/abs/1507.06273} {arXiv:1507.06273 [nucl-ex]} \BibitemShut
  {NoStop}%
%%CITATION = ARXIV:1507.06273;%%
\bibitem [{\citenamefont {Khachatryan}\ \emph {et~al.}(2010)\citenamefont
  {Khachatryan} \emph {et~al.}}]{Khachatryan:2010gv}%
  \BibitemOpen
  \bibfield  {author} {\bibinfo {author} {\bibfnamefont {V.}~\bibnamefont
  {Khachatryan}} \emph {et~al.} (\bibinfo {collaboration} {CMS}),\ }\href
  {\doibase 10.1007/JHEP09(2010)091} {\bibfield  {journal} {\bibinfo  {journal}
  {JHEP}\ }\textbf {\bibinfo {volume} {09}},\ \bibinfo {pages} {091} (\bibinfo
  {year} {2010})},\ \Eprint {http://arxiv.org/abs/1009.4122} {arXiv:1009.4122
  [hep-ex]} \BibitemShut {NoStop}%
%%CITATION = ARXIV:1009.4122;%%
\bibitem [{\citenamefont {Bo\.zek}(2012)}]{Bozek:2011if}%
  \BibitemOpen
  \bibfield  {author} {\bibinfo {author} {\bibfnamefont {P.}~\bibnamefont
  {Bo\.zek}},\ }\href@noop {} {\bibfield  {journal} {\bibinfo  {journal} {Phys.
  Rev.}\ }\textbf {\bibinfo {volume} {C85}},\ \bibinfo {pages} {014911}
  (\bibinfo {year} {2012})},\ \Eprint {http://arxiv.org/abs/1112.0915}
  {arXiv:1112.0915 [hep-ph]} \BibitemShut {NoStop}%
%%CITATION = ARXIV:1112.0915;%%
\bibitem [{\citenamefont {Kozlov}\ \emph {et~al.}(2014)\citenamefont {Kozlov},
  \citenamefont {Luzum}, \citenamefont {Denicol}, \citenamefont {Jeon},\ and\
  \citenamefont {Gale}}]{Kozlov:2014fqa}%
  \BibitemOpen
  \bibfield  {author} {\bibinfo {author} {\bibfnamefont {I.}~\bibnamefont
  {Kozlov}}, \bibinfo {author} {\bibfnamefont {M.}~\bibnamefont {Luzum}},
  \bibinfo {author} {\bibfnamefont {G.}~\bibnamefont {Denicol}}, \bibinfo
  {author} {\bibfnamefont {S.}~\bibnamefont {Jeon}}, \ and\ \bibinfo {author}
  {\bibfnamefont {C.}~\bibnamefont {Gale}},\ }\href@noop {} {\  (\bibinfo
  {year} {2014})},\ \Eprint {http://arxiv.org/abs/1405.3976} {arXiv:1405.3976
  [nucl-th]} \BibitemShut {NoStop}%
%%CITATION = ARXIV:1405.3976;%%
\bibitem [{\citenamefont {Bzdak}\ and\ \citenamefont
  {Ma}(2014)}]{Bzdak:2014dia}%
  \BibitemOpen
  \bibfield  {author} {\bibinfo {author} {\bibfnamefont {A.}~\bibnamefont
  {Bzdak}}\ and\ \bibinfo {author} {\bibfnamefont {G.-L.}\ \bibnamefont {Ma}},\
  }\href {\doibase 10.1103/PhysRevLett.113.252301} {\bibfield  {journal}
  {\bibinfo  {journal} {Phys.Rev.Lett.}\ }\textbf {\bibinfo {volume} {113}},\
  \bibinfo {pages} {252301} (\bibinfo {year} {2014})},\ \Eprint
  {http://arxiv.org/abs/1406.2804} {arXiv:1406.2804 [hep-ph]} \BibitemShut
  {NoStop}%
%%CITATION = ARXIV:1406.2804;%%
\bibitem [{\citenamefont {Bo\.zek}\ and\ \citenamefont
  {Broniowski}(2013)}]{Bozek:2013uha}%
  \BibitemOpen
  \bibfield  {author} {\bibinfo {author} {\bibfnamefont {P.}~\bibnamefont
  {Bo\.zek}}\ and\ \bibinfo {author} {\bibfnamefont {W.}~\bibnamefont
  {Broniowski}},\ }\href {\doibase 10.1103/PhysRevC.88.014903} {\bibfield
  {journal} {\bibinfo  {journal} {Phys. Rev.}\ }\textbf {\bibinfo {volume}
  {C88}},\ \bibinfo {pages} {014903} (\bibinfo {year} {2013})},\ \Eprint
  {http://arxiv.org/abs/1304.3044} {arXiv:1304.3044 [nucl-th]} \BibitemShut
  {NoStop}%
%%CITATION = ARXIV:1304.3044;%%
\bibitem [{\citenamefont {Werner}\ \emph {et~al.}(2014)\citenamefont {Werner},
  \citenamefont {Guiot}, \citenamefont {Karpenko},\ and\ \citenamefont
  {Pierog}}]{Werner:2013tya}%
  \BibitemOpen
  \bibfield  {author} {\bibinfo {author} {\bibfnamefont {K.}~\bibnamefont
  {Werner}}, \bibinfo {author} {\bibfnamefont {B.}~\bibnamefont {Guiot}},
  \bibinfo {author} {\bibfnamefont {I.}~\bibnamefont {Karpenko}}, \ and\
  \bibinfo {author} {\bibfnamefont {T.}~\bibnamefont {Pierog}},\ }\href
  {\doibase 10.1103/PhysRevC.89.064903} {\bibfield  {journal} {\bibinfo
  {journal} {Phys. Rev.}\ }\textbf {\bibinfo {volume} {C89}},\ \bibinfo {pages}
  {064903} (\bibinfo {year} {2014})},\ \Eprint {http://arxiv.org/abs/1312.1233}
  {arXiv:1312.1233 [nucl-th]} \BibitemShut {NoStop}%
%%CITATION = ARXIV:1312.1233;%%
\bibitem [{\citenamefont {Nagle}\ \emph {et~al.}(2014)\citenamefont {Nagle},
  \citenamefont {Adare}, \citenamefont {Beckman}, \citenamefont {Koblesky},
  \citenamefont {Koop} \emph {et~al.}}]{Nagle:2013lja}%
  \BibitemOpen
  \bibfield  {author} {\bibinfo {author} {\bibfnamefont {J.}~\bibnamefont
  {Nagle}}, \bibinfo {author} {\bibfnamefont {A.}~\bibnamefont {Adare}},
  \bibinfo {author} {\bibfnamefont {S.}~\bibnamefont {Beckman}}, \bibinfo
  {author} {\bibfnamefont {T.}~\bibnamefont {Koblesky}}, \bibinfo {author}
  {\bibfnamefont {J.~O.}\ \bibnamefont {Koop}},  \emph {et~al.},\ }\href
  {\doibase 10.1103/PhysRevLett.113.112301} {\bibfield  {journal} {\bibinfo
  {journal} {Phys.Rev.Lett.}\ }\textbf {\bibinfo {volume} {113}},\ \bibinfo
  {pages} {112301} (\bibinfo {year} {2014})},\ \Eprint
  {http://arxiv.org/abs/1312.4565} {arXiv:1312.4565 [nucl-th]} \BibitemShut
  {NoStop}%
%%CITATION = ARXIV:1312.4565;%%
\bibitem [{\citenamefont {Bzdak}\ \emph {et~al.}(2013)\citenamefont {Bzdak},
  \citenamefont {Schenke}, \citenamefont {Tribedy},\ and\ \citenamefont
  {Venugopalan}}]{Bzdak:2013zma}%
  \BibitemOpen
  \bibfield  {author} {\bibinfo {author} {\bibfnamefont {A.}~\bibnamefont
  {Bzdak}}, \bibinfo {author} {\bibfnamefont {B.}~\bibnamefont {Schenke}},
  \bibinfo {author} {\bibfnamefont {P.}~\bibnamefont {Tribedy}}, \ and\
  \bibinfo {author} {\bibfnamefont {R.}~\bibnamefont {Venugopalan}},\ }\href
  {\doibase 10.1103/PhysRevC.87.064906} {\bibfield  {journal} {\bibinfo
  {journal} {Phys. Rev.}\ }\textbf {\bibinfo {volume} {C87}},\ \bibinfo {pages}
  {064906} (\bibinfo {year} {2013})},\ \Eprint {http://arxiv.org/abs/1304.3403}
  {arXiv:1304.3403 [nucl-th]} \BibitemShut {NoStop}%
%%CITATION = ARXIV:1304.3403;%%
\bibitem [{\citenamefont {Broniowski}\ and\ \citenamefont
  {Arriola}(2014)}]{Broniowski:2013dia}%
  \BibitemOpen
  \bibfield  {author} {\bibinfo {author} {\bibfnamefont {W.}~\bibnamefont
  {Broniowski}}\ and\ \bibinfo {author} {\bibfnamefont {E.~R.}\ \bibnamefont
  {Arriola}},\ }\href {\doibase 10.1103/PhysRevLett.112.112501} {\bibfield
  {journal} {\bibinfo  {journal} {Phys. Rev. Lett.}\ }\textbf {\bibinfo
  {volume} {112}},\ \bibinfo {pages} {112501} (\bibinfo {year} {2014})},\
  \Eprint {http://arxiv.org/abs/1312.0289} {arXiv:1312.0289 [nucl-th]}
  \BibitemShut {NoStop}%
%%CITATION = ARXIV:1312.0289;%%
\bibitem [{\citenamefont {Bo{\.z}ek}\ and\ \citenamefont
  {Broniowski}(2014)}]{Bozek:2014cya}%
  \BibitemOpen
  \bibfield  {author} {\bibinfo {author} {\bibfnamefont {P.}~\bibnamefont
  {Bo{\.z}ek}}\ and\ \bibinfo {author} {\bibfnamefont {W.}~\bibnamefont
  {Broniowski}},\ }\href {\doibase 10.1016/j.physletb.2014.11.006} {\bibfield
  {journal} {\bibinfo  {journal} {Phys.Lett.}\ }\textbf {\bibinfo {volume}
  {B739}},\ \bibinfo {pages} {308} (\bibinfo {year} {2014})},\ \Eprint
  {http://arxiv.org/abs/1409.2160} {arXiv:1409.2160 [nucl-th]} \BibitemShut
  {NoStop}%
%%CITATION = ARXIV:1409.2160;%%
\bibitem [{\citenamefont {Aidala}\ \emph {et~al.}(2018)\citenamefont {Aidala}
  \emph {et~al.}}]{Aidala:2018mcw}%
  \BibitemOpen
  \bibfield  {author} {\bibinfo {author} {\bibfnamefont {C.}~\bibnamefont
  {Aidala}} \emph {et~al.} (\bibinfo {collaboration} {PHENIX Collaboration}),\
  }\href@noop {} {\  (\bibinfo {year} {2018})},\ \Eprint
  {http://arxiv.org/abs/1805.02973} {arXiv:1805.02973 [nucl-ex]} \BibitemShut
  {NoStop}%
%%CITATION = ARXIV:1805.02973;%%
\bibitem [{\citenamefont {Dumitru}\ \emph {et~al.}(2008)\citenamefont
  {Dumitru}, \citenamefont {Gelis}, \citenamefont {McLerran},\ and\
  \citenamefont {Venugopalan}}]{Dumitru:2008wn}%
  \BibitemOpen
  \bibfield  {author} {\bibinfo {author} {\bibfnamefont {A.}~\bibnamefont
  {Dumitru}}, \bibinfo {author} {\bibfnamefont {F.}~\bibnamefont {Gelis}},
  \bibinfo {author} {\bibfnamefont {L.}~\bibnamefont {McLerran}}, \ and\
  \bibinfo {author} {\bibfnamefont {R.}~\bibnamefont {Venugopalan}},\ }\href
  {\doibase 10.1016/j.nuclphysa.2008.06.012} {\bibfield  {journal} {\bibinfo
  {journal} {Nucl. Phys.}\ }\textbf {\bibinfo {volume} {A810}},\ \bibinfo
  {pages} {91} (\bibinfo {year} {2008})},\ \Eprint
  {http://arxiv.org/abs/0804.3858} {arXiv:0804.3858 [hep-ph]} \BibitemShut
  {NoStop}%
%%CITATION = ARXIV:0804.3858;%%
\bibitem [{\citenamefont {Dusling}\ and\ \citenamefont
  {Venugopalan}(2012)}]{Dusling:2012iga}%
  \BibitemOpen
  \bibfield  {author} {\bibinfo {author} {\bibfnamefont {K.}~\bibnamefont
  {Dusling}}\ and\ \bibinfo {author} {\bibfnamefont {R.}~\bibnamefont
  {Venugopalan}},\ }\href {\doibase 10.1103/PhysRevLett.108.262001} {\bibfield
  {journal} {\bibinfo  {journal} {Phys. Rev. Lett.}\ }\textbf {\bibinfo
  {volume} {108}},\ \bibinfo {pages} {262001} (\bibinfo {year} {2012})},\
  \Eprint {http://arxiv.org/abs/1201.2658} {arXiv:1201.2658 [hep-ph]}
  \BibitemShut {NoStop}%
%%CITATION = ARXIV:1201.2658;%%
\bibitem [{\citenamefont {Kovchegov}\ and\ \citenamefont
  {Wertepny}(2013)}]{Kovchegov:2012nd}%
  \BibitemOpen
  \bibfield  {author} {\bibinfo {author} {\bibfnamefont {Y.~V.}\ \bibnamefont
  {Kovchegov}}\ and\ \bibinfo {author} {\bibfnamefont {D.~E.}\ \bibnamefont
  {Wertepny}},\ }\href {\doibase 10.1016/j.nuclphysa.2013.03.006} {\bibfield
  {journal} {\bibinfo  {journal} {Nucl.Phys.}\ }\textbf {\bibinfo {volume}
  {A906}},\ \bibinfo {pages} {50} (\bibinfo {year} {2013})},\ \Eprint
  {http://arxiv.org/abs/1212.1195} {arXiv:1212.1195 [hep-ph]} \BibitemShut
  {NoStop}%
%%CITATION = ARXIV:1212.1195;%%
\bibitem [{\citenamefont {Dusling}\ and\ \citenamefont
  {Venugopalan}(2013)}]{Dusling:2013oia}%
  \BibitemOpen
  \bibfield  {author} {\bibinfo {author} {\bibfnamefont {K.}~\bibnamefont
  {Dusling}}\ and\ \bibinfo {author} {\bibfnamefont {R.}~\bibnamefont
  {Venugopalan}},\ }\href {\doibase 10.1103/PhysRevD.87.094034} {\bibfield
  {journal} {\bibinfo  {journal} {Phys. Rev.}\ }\textbf {\bibinfo {volume}
  {D87}},\ \bibinfo {pages} {094034} (\bibinfo {year} {2013})},\ \Eprint
  {http://arxiv.org/abs/1302.7018} {arXiv:1302.7018 [hep-ph]} \BibitemShut
  {NoStop}%
%%CITATION = ARXIV:1302.7018;%%
\bibitem [{\citenamefont {Mace}\ \emph
  {et~al.}(2018{\natexlab{a}})\citenamefont {Mace}, \citenamefont {Skokov},
  \citenamefont {Tribedy},\ and\ \citenamefont {Venugopalan}}]{Mace:2018vwq}%
  \BibitemOpen
  \bibfield  {author} {\bibinfo {author} {\bibfnamefont {M.}~\bibnamefont
  {Mace}}, \bibinfo {author} {\bibfnamefont {V.~V.}\ \bibnamefont {Skokov}},
  \bibinfo {author} {\bibfnamefont {P.}~\bibnamefont {Tribedy}}, \ and\
  \bibinfo {author} {\bibfnamefont {R.}~\bibnamefont {Venugopalan}},\ }\href
  {\doibase 10.1103/PhysRevLett.121.052301} {\bibfield  {journal} {\bibinfo
  {journal} {Phys. Rev. Lett.}\ }\textbf {\bibinfo {volume} {121}},\ \bibinfo
  {pages} {052301} (\bibinfo {year} {2018}{\natexlab{a}})},\ \Eprint
  {http://arxiv.org/abs/1805.09342} {arXiv:1805.09342 [hep-ph]} \BibitemShut
  {NoStop}%
%%CITATION = ARXIV:1805.09342;%%
\bibitem [{\citenamefont {Mace}\ \emph
  {et~al.}(2018{\natexlab{b}})\citenamefont {Mace}, \citenamefont {Skokov},
  \citenamefont {Tribedy},\ and\ \citenamefont {Venugopalan}}]{Mace:2018yvl}%
  \BibitemOpen
  \bibfield  {author} {\bibinfo {author} {\bibfnamefont {M.}~\bibnamefont
  {Mace}}, \bibinfo {author} {\bibfnamefont {V.~V.}\ \bibnamefont {Skokov}},
  \bibinfo {author} {\bibfnamefont {P.}~\bibnamefont {Tribedy}}, \ and\
  \bibinfo {author} {\bibfnamefont {R.}~\bibnamefont {Venugopalan}},\
  }\href@noop {} {\  (\bibinfo {year} {2018}{\natexlab{b}})},\ \Eprint
  {http://arxiv.org/abs/1807.00825} {arXiv:1807.00825 [hep-ph]} \BibitemShut
  {NoStop}%
%%CITATION = ARXIV:1807.00825;%%
\bibitem [{\citenamefont {Zhaba}(2016)}]{Zhaba:2015yxq}%
  \BibitemOpen
  \bibfield  {author} {\bibinfo {author} {\bibfnamefont {V.~I.}\ \bibnamefont
  {Zhaba}},\ }\href {\doibase 10.1142/S021773231650139X} {\bibfield  {journal}
  {\bibinfo  {journal} {Mod. Phys. Lett.}\ }\textbf {\bibinfo {volume} {A31}},\
  \bibinfo {pages} {1650139} (\bibinfo {year} {2016})},\ \Eprint
  {http://arxiv.org/abs/1512.08980} {arXiv:1512.08980 [nucl-th]} \BibitemShut
  {NoStop}%
%%CITATION = ARXIV:1512.08980;%%
\bibitem [{\citenamefont {Bia\l{}as}\ \emph {et~al.}(1976)\citenamefont
  {Bia\l{}as}, \citenamefont {B\l{}eszy\'nski},\ and\ \citenamefont
  {Czy\.z}}]{Bialas:1976ed}%
  \BibitemOpen
  \bibfield  {author} {\bibinfo {author} {\bibfnamefont {A.}~\bibnamefont
  {Bia\l{}as}}, \bibinfo {author} {\bibfnamefont {M.}~\bibnamefont
  {B\l{}eszy\'nski}}, \ and\ \bibinfo {author} {\bibfnamefont {W.}~\bibnamefont
  {Czy\.z}},\ }\href@noop {} {\bibfield  {journal} {\bibinfo  {journal} {Nucl.
  Phys.}\ }\textbf {\bibinfo {volume} {B111}},\ \bibinfo {pages} {461}
  (\bibinfo {year} {1976})}\BibitemShut {NoStop}%
%%CITATION = NUPHA,B111,461;%%
\bibitem [{\citenamefont {Kharzeev}\ and\ \citenamefont
  {Nardi}(2001)}]{Kharzeev:2000ph}%
  \BibitemOpen
  \bibfield  {author} {\bibinfo {author} {\bibfnamefont {D.}~\bibnamefont
  {Kharzeev}}\ and\ \bibinfo {author} {\bibfnamefont {M.}~\bibnamefont
  {Nardi}},\ }\href {\doibase 10.1016/S0370-2693(01)00457-9} {\bibfield
  {journal} {\bibinfo  {journal} {Phys. Lett.}\ }\textbf {\bibinfo {volume}
  {B507}},\ \bibinfo {pages} {121} (\bibinfo {year} {2001})},\ \Eprint
  {http://arxiv.org/abs/nucl-th/0012025} {arXiv:nucl-th/0012025} \BibitemShut
  {NoStop}%
%%CITATION = NUCL-TH/0012025;%%
\bibitem [{\citenamefont {Rybczy\'nski}\ \emph {et~al.}(2014)\citenamefont
  {Rybczy\'nski}, \citenamefont {Stefanek}, \citenamefont {Broniowski},\ and\
  \citenamefont {Bo\.zek}}]{Rybczynski:2013yba}%
  \BibitemOpen
  \bibfield  {author} {\bibinfo {author} {\bibfnamefont {M.}~\bibnamefont
  {Rybczy\'nski}}, \bibinfo {author} {\bibfnamefont {G.}~\bibnamefont
  {Stefanek}}, \bibinfo {author} {\bibfnamefont {W.}~\bibnamefont
  {Broniowski}}, \ and\ \bibinfo {author} {\bibfnamefont {P.}~\bibnamefont
  {Bo\.zek}},\ }\href {\doibase 10.1016/j.cpc.2014.02.016} {\bibfield
  {journal} {\bibinfo  {journal} {Comput. Phys. Commun.}\ }\textbf {\bibinfo
  {volume} {185}},\ \bibinfo {pages} {1759} (\bibinfo {year} {2014})},\ \Eprint
  {http://arxiv.org/abs/1310.5475} {arXiv:1310.5475 [nucl-th]} \BibitemShut
  {NoStop}%
%%CITATION = ARXIV:1310.5475;%%
\bibitem [{\citenamefont {Mane}\ \emph {et~al.}(2005)\citenamefont {Mane},
  \citenamefont {Shatunov},\ and\ \citenamefont {Yokoya}}]{Mane:2005xh}%
  \BibitemOpen
  \bibfield  {author} {\bibinfo {author} {\bibfnamefont {S.~R.}\ \bibnamefont
  {Mane}}, \bibinfo {author} {\bibfnamefont {{\relax Yu}.~M.}\ \bibnamefont
  {Shatunov}}, \ and\ \bibinfo {author} {\bibfnamefont {K.}~\bibnamefont
  {Yokoya}},\ }\href {\doibase 10.1088/0034-4885/68/9/R01} {\bibfield
  {journal} {\bibinfo  {journal} {Rept. Prog. Phys.}\ }\textbf {\bibinfo
  {volume} {68}},\ \bibinfo {pages} {1997} (\bibinfo {year}
  {2005})}\BibitemShut {NoStop}%
%%CITATION = RPPHA,68,1997;%%
\bibitem [{\citenamefont {Alekseev}\ \emph {et~al.}(2003)\citenamefont
  {Alekseev} \emph {et~al.}}]{Alekseev:2003sk}%
  \BibitemOpen
  \bibfield  {author} {\bibinfo {author} {\bibfnamefont {I.}~\bibnamefont
  {Alekseev}} \emph {et~al.},\ }\href {\doibase 10.1016/S0168-9002(02)01946-0}
  {\bibfield  {journal} {\bibinfo  {journal} {Nucl. Instrum. Meth.}\ }\textbf
  {\bibinfo {volume} {A499}},\ \bibinfo {pages} {392} (\bibinfo {year}
  {2003})}\BibitemShut {NoStop}%
%%CITATION = NUIMA,A499,392;%%
\bibitem [{\citenamefont {Sato}\ \emph {et~al.}(1997)\citenamefont {Sato} \emph
  {et~al.}}]{Sato:1996te}%
  \BibitemOpen
  \bibfield  {author} {\bibinfo {author} {\bibfnamefont {H.}~\bibnamefont
  {Sato}} \emph {et~al.},\ }\href {\doibase 10.1016/S0168-9002(96)01169-2}
  {\bibfield  {journal} {\bibinfo  {journal} {Nucl. Instrum. Meth.}\ }\textbf
  {\bibinfo {volume} {A385}},\ \bibinfo {pages} {391} (\bibinfo {year}
  {1997})}\BibitemShut {NoStop}%
%%CITATION = NUIMA,A385,391;%%
\bibitem [{\citenamefont {Savin}\ \emph {et~al.}(2015)\citenamefont {Savin}
  \emph {et~al.}}]{Savin:2014sva}%
  \BibitemOpen
  \bibfield  {author} {\bibinfo {author} {\bibfnamefont {I.~A.}\ \bibnamefont
  {Savin}} \emph {et~al.},\ }\bibfield  {booktitle} {\emph {\bibinfo
  {booktitle} {{Proceedings, 4th International Workshop on Transverse
  Polarization Phenomena in Hard Processes (Transversity 2014): Cagliari,
  Italy, June 9-13, 2014}}},\ }\href {\doibase 10.1051/epjconf/20158502039}
  {\bibfield  {journal} {\bibinfo  {journal} {EPJ Web Conf.}\ }\textbf
  {\bibinfo {volume} {85}},\ \bibinfo {pages} {02039} (\bibinfo {year}
  {2015})},\ \Eprint {http://arxiv.org/abs/1408.3959} {arXiv:1408.3959
  [hep-ex]} \BibitemShut {NoStop}%
%%CITATION = ARXIV:1408.3959;%%
\bibitem [{\citenamefont {Borghini}\ \emph {et~al.}(2001)\citenamefont
  {Borghini}, \citenamefont {Dinh},\ and\ \citenamefont
  {Ollitrault}}]{Borghini:2001vi}%
  \BibitemOpen
  \bibfield  {author} {\bibinfo {author} {\bibfnamefont {N.}~\bibnamefont
  {Borghini}}, \bibinfo {author} {\bibfnamefont {P.~M.}\ \bibnamefont {Dinh}},
  \ and\ \bibinfo {author} {\bibfnamefont {J.-Y.}\ \bibnamefont {Ollitrault}},\
  }\href {\doibase 10.1103/PhysRevC.64.054901} {\bibfield  {journal} {\bibinfo
  {journal} {Phys.Rev.}\ }\textbf {\bibinfo {volume} {C64}},\ \bibinfo {pages}
  {054901} (\bibinfo {year} {2001})},\ \Eprint
  {http://arxiv.org/abs/nucl-th/0105040} {arXiv:nucl-th/0105040 [nucl-th]}
  \BibitemShut {NoStop}%
%%CITATION = NUCL-TH/0105040;%%
\bibitem [{\citenamefont {Abgrall}\ \emph {et~al.}(2014)\citenamefont {Abgrall}
  \emph {et~al.}}]{Abgrall:2014xwa}%
  \BibitemOpen
  \bibfield  {author} {\bibinfo {author} {\bibfnamefont {N.}~\bibnamefont
  {Abgrall}} \emph {et~al.} (\bibinfo {collaboration} {NA61}),\ }\href
  {\doibase 10.1088/1748-0221/9/06/P06005} {\bibfield  {journal} {\bibinfo
  {journal} {JINST}\ }\textbf {\bibinfo {volume} {9}},\ \bibinfo {pages}
  {P06005} (\bibinfo {year} {2014})},\ \Eprint {http://arxiv.org/abs/1401.4699}
  {arXiv:1401.4699 [physics.ins-det]} \BibitemShut {NoStop}%
%%CITATION = ARXIV:1401.4699;%%
\bibitem [{\citenamefont {Barschel}(2014)}]{Barschel:2014iua}%
  \BibitemOpen
  \bibfield  {author} {\bibinfo {author} {\bibfnamefont {C.}~\bibnamefont
  {Barschel}},\ }\emph {\bibinfo {title} {{Precision luminosity measurement at
  LHCb with beam-gas imaging}}},\ \href@noop {} {Ph.D. thesis},\ \bibinfo
  {school} {RWTH Aachen U.} (\bibinfo {year} {2014})\BibitemShut {NoStop}%
%%CITATION = CERN-THESIS-2013-301;%%
\bibitem [{\citenamefont {Aaij}\ \emph {et~al.}(2014)\citenamefont {Aaij} \emph
  {et~al.}}]{Aaij:2014ida}%
  \BibitemOpen
  \bibfield  {author} {\bibinfo {author} {\bibfnamefont {R.}~\bibnamefont
  {Aaij}} \emph {et~al.} (\bibinfo {collaboration} {LHCb}),\ }\href {\doibase
  10.1088/1748-0221/9/12/P12005} {\bibfield  {journal} {\bibinfo  {journal}
  {JINST}\ }\textbf {\bibinfo {volume} {9}},\ \bibinfo {pages} {P12005}
  (\bibinfo {year} {2014})},\ \Eprint {http://arxiv.org/abs/1410.0149}
  {arXiv:1410.0149 [hep-ex]} \BibitemShut {NoStop}%
%%CITATION = ARXIV:1410.0149;%%
\end{thebibliography}%

\end{document}